\documentclass[twocolumn,showpacs,prl,10pt]{revtex4}
\usepackage{graphicx}

\begin{document}

\title {On the nonlinear response of a particle interacting\\ with 
fermions in a 1D lattice}
\author{X. Zotos}
\affiliation{Department of Physics, University of Crete\\ and\\
Foundation for Research and Technology-Hellas, P.O. Box 2208, 71003
Heraklion, Greece}

\date{\today}

\begin{abstract}
By the Bethe ansatz method we study the energy dispersion of a particle 
interacting by a local interaction with fermions (or hard core bosons) 
of equal mass in a one dimensional lattice. We focus on the period of the 
Bloch oscillations which turns out to be related to the Fermi wavevector 
of the Fermi sea and in particular on how this dispersion emerges as 
a collective effect in the thermodynamic limit. We show 
by symmetry that the dispersion is temperature independent for a 
half-filled system. We also discuss the adiabatic coherent collective  
response of the particle to an applied field.
\end{abstract}

\maketitle

Prototype integrable quantum many-body problems provide valuable insight into 
the physics of correlated electronic and magnetic systems. 
With the development of novel experimental systems, 
as quasi-one dimensional quantum magnets \cite{hess} and 
cold atom systems \cite{cold}, it became possible to tailor make 
and experimentally study these systems. At the same time because of their 
nongeneric character they exhibit unconventional dynamic and transport 
properties of academic and potentially technological interest.

One of the simplest models that was studied early on \cite{mcguire}, 
is that of a particle interacting with spinless fermions in a one 
dimensional lattice, where the mass of the extra particle is the same 
as the mass of the fermions. In this model 
and its lattice version, 
it was shown  that, despite the particle-fermion interaction, 
the transport of the ''heavy" particle remains ballistic at all 
temperatures, namely its mobility diverges \cite{xz,czp}.
The spinless fermion or hard core boson bath corresponds to the 
Tonks-Girardeau gas, a one dimensional boson model with very strong 
local repulsion.

More recently \cite{gangardt, lamacraft, girardeau}, motivated by 
cold atom physics experiments, 
the dispersion and period of Bloch oscillations of the heavy particle when 
acted by a constant field in a continuous system were also studied 
and shown to be related to the Fermi wavevector $k_F$ of the 
fermionic sea. 
The first issue we study in this work is the competition between 
the period of oscillations imposed by the lattice in the tight binding 
version of the model and that imposed by the Fermi wavevector.

The lattice model corresponds to a 
particular sector of the 1D Hubbard model \cite{czp}, 
a prototype integrable model by the Bethe ansatz method \cite{lieb}. 
The analytical solution allows us to characterize the low energy spectrum 
and also study the adiabatic response 
of the particle to an applied field.

The system is described by the Hamiltonian, 
\begin{eqnarray*}
\hat H&=&-t_h\sum_l (e^{i\phi} d^{\dagger}_{l+1} d_l +h.c.)
-t\sum_l (c^{\dagger}_{l+1} c_l + h.c.)\nonumber\\  
&+& U\sum_l d^{\dagger}_l d_l c^{\dagger}_l c_l,\nonumber
\end{eqnarray*}
where  $c_l ( c^{\dagger}_l )$ are annihilation (creation) operators for
$N$ spinless fermions and $d_l ( d^{\dagger}_l )$ of the extra particle
on an $L$ site chain with periodic boundary conditions. The interaction
comes only through the on-site repulsion $U>0$.
If the particle is charged and $\phi$ depends linearly on time,  
$\phi=-{\cal E} t$, then a constant electric field 
${\cal E}=-\partial \phi/\partial t$ acts on it. 
For a vanishing field 
(it could also be gravitational \cite{cold}) the system should follow the 
adiabatic ground state that we will now map.

\begin{figure}[ht]
\includegraphics[angle=0, width=1.0\linewidth]{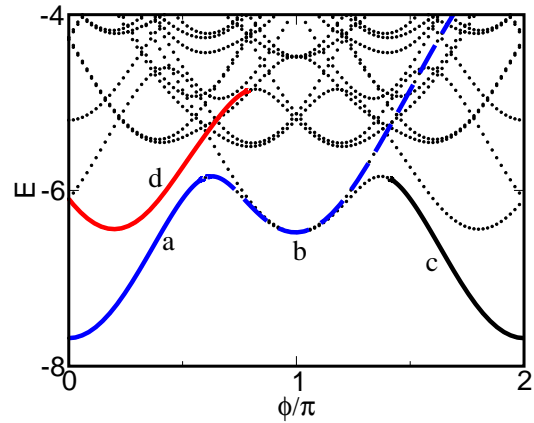}
\caption{Low energy spectrum for $L=10,~~u=0.5, t_h=t$, 
along with the reconsruction 
of selected branches by the Bethe ansatz solution.}
\label{fig1}
\end{figure}

For illustration of the typical finite size lattice spectra, we show  
in Fig. \ref{fig1} the evolution of the low lying energy levels for $L=10$, 
half- filling  and in the $k=0$ symmetry momentum subspace 
as a function of $\phi$.
It is clear that by scanning $\phi$ we recover at $\phi=2\pi n/L$ ($n=0,L-1$) 
the spectra in the succesive $k-$subspaces with $\phi=0$. 
We will now show how to reconstruct selected branches of the dispersion 
and study their finite size effects using the Bethe ansatz solution.

\bigskip
The Bethe ansatz wavefunctions, in the presence of flux $\phi$,
are characterized by $M=N+1$ quantum numbers $k_j$ for $N$ fermions plus the 
one extra particle, given
by the following equations (we take $t_h=t$),
\begin{equation}
k_j=\frac{2\pi I_j}{L}+\frac{1}{L}\theta(\sin k_j-\Lambda),~~~j=1,\ldots,M 
\label{kj}
\end{equation}
\begin{equation}
\theta(p)=-2\tan^{-1}(p/u),~~~u=U/4t 
\end{equation}
\begin{eqnarray}
K=\sum_{j=1}^M k_j&=&\sum_{j=1}^M \frac{2\pi I_j}{L}+\frac{2\pi J}{L} + \phi
\nonumber\\
&=&\sum_{j=1}^M  k_j^0 +\frac{2\pi J}{L} + \phi.
\label{k}
\end{eqnarray}
\noindent
The total energy and momentum are given by the sum of quasi-energies 
and momenta,
\begin{equation}
E=\sum_{j=1}^M \epsilon(k_j)=
-2t\sum_{j=1}^M \cos k_j ,~~~K=\sum_{j=1}^M k_j.
\label{ek}
\end{equation}
Every state is characterized by a set of half-odd integers $I_j$
and the (half-odd) integer $J$ for (even) odd number of fermions.
We will consider the case of odd number of fermions, $M$ even, that is also 
equivalent to that of hard core bosons in a one dimensional system.

We can reconstruct the low energy states by considering the following
branches;

\noindent
(a) place the $I_j's$ at the 
successive values $ -M/2+1/2  \le I_j \le +M/2-1/2$ and scan $\Lambda$ 
between $\pm \infty$.  
At this configuration, for $u \rightarrow 0$ and $\Lambda=0$ the phase 
shift term in eq.(\ref{kj}) 
takes the value $+\pi/L (-\pi/L)$ for $k_j <0 (k_j >0)$. 
Thus for one particle $k_j=0$ while the rest of fermions fill 
uniformly a Fermi sea, between $ -(N-1)/2 \le k_j \le +(N-1)/2 $. 
The k-occupation corresponds to that of independent particles.

For a finite $u$, by varying $\Lambda$ every $k_j$ 
shifts by $-\pi/L (+\pi/L) $ for $\Lambda \rightarrow -\infty (+\infty) $  
so that $ -2M\pi/L \le K \le  +2M\pi/L$. In the thermodynamic limit 
$ -k_F \le  K \le +k_F$ with $k_F=\pi N/L$. It is clear that the 
spectrum is scanned by finding the $\Lambda$ solutions of eqs.(\ref{kj})
as a function of $\phi$ for $J=0$ or equivalently for $\phi=0$ as a function 
of the quantum number $J$. 

\noindent
(b) shift each $I_j$ of branch (a) by +1. Now by varying 
$-\infty < \Lambda < +\infty$, $ \pi M/L \le K \le 3\pi M/L $ 
which tends in the thermodynamic limit to $k_F \le K \le 3k_F$. 

\noindent
(c) shift any $I_j$ of branch (a) by $L$ and scan $\Lambda$.

\noindent
(d) shift $I_M$ of branch (a) by $+1$ and scan $\Lambda$.

The adiabatically evolving ground state is obtained by the successive shift 
of $I_j \rightarrow I_j+1$ up to $L$ times when the $k_j's$ are shifted 
by the trivial phase $2\pi$.
Branch (d) corresponds to a state with an ''electron-hole" excitation 
that is not adiabatically evolving from the ground state.

\begin{figure}[ht]
\includegraphics[angle=0, width=1.0\linewidth]{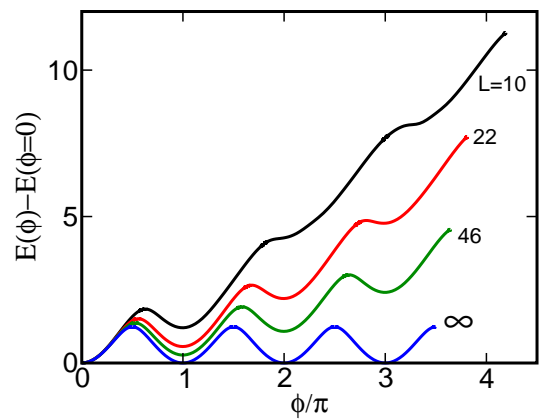}
\caption{Finite size scaling of adiabatically evolving ground state for 
$t_h=t,~~u=0.5$ and half-filling.}
\label{fig2}
\end{figure}

Next, in Fig. \ref{fig2} we show the finite size dependence of the 
adiabatially evolving ground state for a series of lattices at 
half-filling ($k_F=\pi/2$).
It is now clear that in the thermodynamic limit the period 
becomes $2k_F$. 
In order to analyze the finite size effects and obtain the particle dispersion 
we take to $O(1/L)$,
\begin{equation}
k_j\simeq k_j^0+\frac{1}{L}\theta(\sin k_j^0-\Lambda),~~~j=1,\ldots,M. 
\label{kj0}
\end{equation}
We can now see how the dispersion becomes symmetric around $k_F$ 
in the thermodynamic limit by evaluating the difference in energy 
between the state $K=0$ and $K=2k_F$. By shifting $k_1=(2\pi/L)(-M/2+1/2)$ to 
$k_{1'}=(2\pi/L)(M/2+1/2)$ and taking $\Lambda=0$, we obtain,
\begin{equation}  
K_{2k_F}-K_0=\frac{2\pi}{L}M+O(1/L)\simeq 2k_F\nonumber\\
\end{equation}  
\begin{equation}
E_{2k_F}-E_{0}=+2t\cos(k_1)-2t\cos(k_{1'})\simeq \frac{4\pi}{L}\sin{k_F}.
\nonumber\\
\end{equation}
So the finite size effects decrease as $1/L$.
Next we can obtain an analytical solution of the dispersion 
of branch (a) and by symmetry (b) in the thermodynamic limit including 
corrections $O(1/L)$ from eqs. (\ref{k},\ref{ek},\ref{kj0}), 
\begin{eqnarray}
K&=&\sum_{j=1}^M k_j\simeq K_0+\delta K\nonumber\\
&=&\sum_{j=1}^M k_j^0  
+\frac{1}{L}\sum_{j=1}^M\theta(\sin k_j^0-\Lambda)
\label{deltak}
\end{eqnarray}
\begin{eqnarray}
E&=&-2t\sum_{j=1}^M \cos k_j\simeq E_0+\delta E
\nonumber\\
&=&-2t\sum_{j=1}^M \cos k_j^0  
+\frac{2t}{L}\sum_{j=1}^M\sin k_j^0\theta(\sin k_j^0-\Lambda)
\label{deltae}
\end{eqnarray}
Replacing sums by integrals for the branch (a) we find for the 
momentum and energy dependendent terms on $\Lambda$ and thus $\phi$,
\begin{eqnarray}
\delta K&=& \frac{1}{2\pi} \int_{-k_F}^{+k_F} dk\theta(\sin k-\Lambda)
\nonumber\\
\delta E&=&\frac{2t}{2\pi}\int_{-k_F}^{+k_F} \sin k\theta(\sin k-\Lambda).
\end{eqnarray}
\noindent
Scanning $-\infty < \Lambda < +\infty$ we obtain 
the dispersion shown in Fig. \ref{fig2} for $L\rightarrow +\infty$.

If we interpret $\delta E, \delta K$ as the dispersion of the correlation 
energy and momentum 
we can deduce their temperature dependence by inserting 
in eqs. (\ref{deltak},\ref{deltae}) a Fermi-Dirac thermal distribution 
$f^0_k$ for the momenta $k_j^0$. 
\begin{eqnarray}
\delta K&=& \frac{1}{2\pi} \int_{-\pi}^{+\pi} dk
f^0_k\theta(\sin k-\Lambda),
\nonumber\\
\delta E&=&\frac{2t}{2\pi}\int_{-\pi}^{+\pi} 
f^0_k \sin k\theta(\sin k-\Lambda),
\nonumber\\
f^0_k&=&\frac{1}{1+e^{(\epsilon(k^0)-\mu)/k_BT}}
\end{eqnarray}
Now, by electron-hole symmetry we see that at half-filling, $\mu=0$, 
the thermal distribution factors cancel so that the dispersion 
becomes temperature independent. This is due to the particular form 
of scattering phase shifts in this integrable model.

The next issue we want to discuss is the response of the particle 
to a constant external field ${\cal E}$ created by a time dependent flux 
$\phi=-{\cal E}t$.
As is evident from Fig. \ref{fig1} the collective ground state 
evolves through a maze of excited ''electron-hole" states by level crossing. 
The coherent evolution throught the energy spectrum is protected by the 
macroscopic number of conservation laws characteristic of an 
integrable system that suppress level repulsion and also 
result to Poisson statistics.  There is a coherent 
drag of the Fermi sea by the ''heavy" particle.
We should note that a similar level 
crossing trajectory also characterizes the time evolution of every 
excited state.
We can plausibly argue that this is a generic behavior of integrable 
quantum many-body systems.
\begin{figure}[ht]
\includegraphics[angle=0, width=1.0\linewidth]{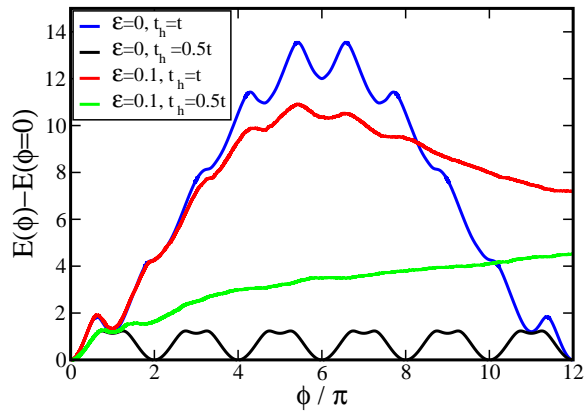}
\caption{Time evolution for $L=10,~~~u=0.5$ at $t_h=t$, 
and $t_h=0.5t$.} 
\label{fig3}
\end{figure}

To put this evolution into evidence 
we show in Fig. \ref{fig3} the long time adiabatic energy evolution 
of an $L=10$ system as a function 
of phase $\phi=-{\cal E} t$ in units of $\pi$  for ${\cal E}\rightarrow 0.$ 
When the momentum of each 
particle is displaced by $2\pi$ so that $\phi=2\pi M$ 
we have an identical state (in the example 
at $\phi/\pi=2M, M=6$). Thus the periodicity of the coherent 
collective motion becomes macroscopic.
In the adiabatic limit there is reversible pumping 
of a macroscopic amount of energy into the system that it could 
experimentally be observed in a mesoscopic system. 
In contrast,  in the 
nonintegrable case e.g. $t_h=0.5t$, the periodicity is simply $2\pi$ for a 
finite size system  and it becomes $2k_F$ in the thermodynamic limit.
This behavior is generic for any interaction $U$ and hopping 
difference $t_h\ne t$.
Note also that in the continuous system, hard core bosons in 1D, the driving 
leads to an indefinite increase of energy as 
there is no $\phi=2\pi M$ periodicity.

As shown in Fig. \ref{fig3}, for a finite field e.g. ${\cal E}=0.1$, 
in the integrable case the energy 
deviates from the adiabatic - nonlinear in energy - trajectory, in sharp 
contrast to the nonadiabatic case where it spreads diffusively upwards 
from the ground state.
It is clear that in the integrable 
system the departure from the adiabatic evolution is by mixing with 
highly excited states 
while in the nonintegrable case by tunneling between 
level repelled nearest neighbor states.
For long times, the integrable system rapidly 
evolves through the whole energy spectrum to thermalize at the infinite 
temperature energy limit (mean value of energy spectrum) in a nonmonotonic way
while in the nonintegrable system diffuses monotonically to the same 
mean energy (not shown). Of course this unconventional thermalization is 
irrelevant in the thermodynamic limit but it might be observable in a 
finite size system.

\begin{figure}[ht]
\includegraphics[angle=0, width=1.0\linewidth]{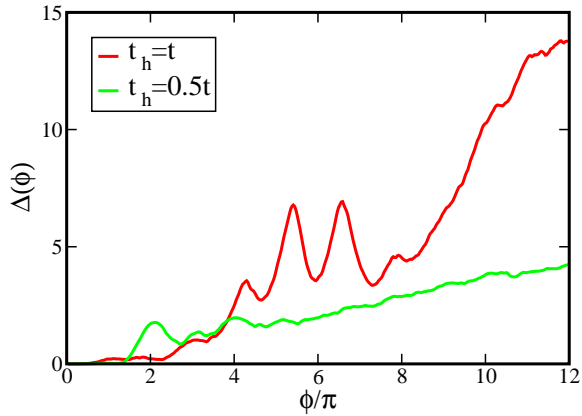}
\caption{Time evolution of wavefunction spread $\Delta(\phi)$ 
for $L=10,~~u=0.5$ at $t_h=t$, 
$t_h=0.5t$ and ${\cal E}=0.1$.}
\label{fig4}
\end{figure}

To quantify the spreading of the wavefunction during the 
evolution, we show in the Fig.\ref{fig4} $\Delta(\phi)$ defined 
by \cite{wilkinson},
\begin{equation}
\Delta(\phi)=<\Psi(\phi)|
(\text{H}(\phi)-<\text{H}(\phi)>)^2|\Psi(\phi)>\nonumber
\end{equation}
Here we also observe that the spreading is highly irregular in the integrable 
case while it is almost linear in time, diffusive, in the nonintegrable one.

The main issues emerging from these results are, first, whether 
it is possible to 
observe experimentally the adiabatic dispersion of the driven particle
in a macroscopic system, that is in fact qualitatively similar in the
integrable and nonintegrable case.
In this context we should mention that recent 
works argue on the absence of adiabatic limit 
in low dimensional gapless systems \cite{gritsev} in macroscopic systems. 
Still, despite the spreading of the wavefunction, it might still be possible 
to observe the characteristic energy-momentum dependence for at least
a few Bloch oscillations and the remnants of the  singular integrable 
behavior. 

Second, in a finite size - mesoscopic - system, there is a striking 
difference between the nonlinear evolution of an integrable and 
a nonintegrable system. 
It is worth studying to what extent this coherent behavior is affected by 
nonadiabatic effects, for at least some modes of driving.
It is experimentally  
interesting to tune the boson-boson interaction from the weak 
(nonintegrable) to the strong Tonks-Girardeau (integrable) limit and 
observe the resulting dynamics. Or alternatively vary the "heavy" particle 
mass.

We conclude that this prototype integrable model exhibits an unconventional 
collective nonlinear evolution when driven by a constant field 
that motivates the exploration of other 
similar integrable models. 

\begin{acknowledgments}
It is a pleasure to thank D. Gangardt and A. Kolovsky for discussions. 
This work was supported by the FP6-032980-2 NOVMAG project. 
\end{acknowledgments}


\begin{references}

\bibitem{hess} C. Hess, 
Eur. Ph. J. Special Topics, {\bf 151}, 73 (2007).

\bibitem{cold} S. Palzer, C. Zipkes, C. Sias and M. K\"ohl,
Phys. Rev. Lett. {\bf 103}, 150601 (2009).

\bibitem{mcguire} J.B. McGuire, J. Math. Phys. {\bf 6}, 432 (1965);
J. Math. Phys. {\bf 7}, 123 (1966).

\bibitem{xz}
X. Zotos, J. Low Temp. Phys. {\bf 126}, 1185 (2002).

\bibitem{czp} C. Castella, X. Zotos and P. Prelov\v sek, 
Phys. Rev. Lett. {\bf 74}, 972 (1995).

\bibitem{gangardt} D.M. Gangardt and A. Kamenev,
Phys. Rev. Lett. {\bf 102}, 070402 (2009).

\bibitem{lamacraft} A. Lamacraft,  
Phys. Rev. B{\bf 79}, 241105(R) (2009).

\bibitem{girardeau} M.D. Girardeau and A. Minguzzi,  
Phys. Rev. A{\bf 79}, 033610 (2009); S. Giraud and R. Combescot, 
arXiv:1002.4366

\bibitem{lieb} E.H. Lieb and F.Y. Wu, Phys. Rev. Lett. {\bf 20}, 1445 (1968).

\bibitem{wilkinson} M. Wilkinson,  Phys. Rev. A{\bf 41}, 4645 (1990).

\bibitem{gritsev} A. Polkovnikov, V. Gritsev,  Nature {\bf 4}, 477 (2008).

\end{references}
\end{document}